\documentclass[
    ,final            
  ]
  {aipproc}

\layoutstyle{8x11single}

\usepackage{amsmath}
\usepackage{color}

\definecolor{Blue}{rgb}{0,0.08,0.65}
\definecolor{Red}{rgb}{0.65,0.08,0.05}
\definecolor{Green}{rgb}{0.15,0.45,0.25}

\begin{document}

\title{The Skeleton: Connecting Large Scale Structures to Galaxy Formation}

\classification{95.35.+d,
95.36.+x  98.80.-k, 98.80.-k
98.80.Jk
}
\keywords      {cosmology, larges scales structures, topology}

\author{Christophe Pichon}{
  address={Department of Physics, Denys Wilkinson Building, Keble Road, Oxford, OX1 3RH, United Kingdom},altaddress={Institut d'Astrophysique de Paris, UPMC Univ Paris 06, CNRS, UMR7095, F-75014, Paris, France}
}

\author{Christophe Gay}{
  address={Institut d'Astrophysique de Paris, UPMC Univ Paris 06, CNRS, UMR7095, F-75014, Paris, France}
}

\author{Dmitry Pogosyan}{
  address={ Department of Physics, University of Alberta, 11322-89 Avenue, Edmonton, Alberta, T6G 2G7, Canada},altaddress={Institut d'Astrophysique de Paris, UPMC Univ Paris 06, CNRS, UMR7095, F-75014, Paris, France}
}

\author{Simon Prunet}{
  address={Institut d'Astrophysique de Paris, UPMC Univ Paris 06, CNRS, UMR7095, F-75014, Paris, France}}

\author{Thierry Sousbie}{
  address={Institut d'Astrophysique de Paris, UPMC Univ Paris 06, CNRS, UMR7095, F-75014, Paris, France},altaddress={Tokyo University, Physics Dept 7-3-1 Hongo Bunkyo-ku, JP Tokyo, 113-0033, Japan}
}

\author{Stephane Colombi}{
  address={Institut d'Astrophysique de Paris, UPMC Univ Paris 06, CNRS, UMR7095, F-75014, Paris, France}
}

\author{Adrianne  Slyz}{
  address={Department of Physics, Denys Wilkinson Building, Keble Road, Oxford, OX1 3RH, United Kingdom}
}

\author{Julien  Devriendt}{
  address={Department of Physics, Denys Wilkinson Building, Keble Road, Oxford, OX1 3RH, United Kingdom}
}

\begin{abstract}
We report on two quantitative, morphological estimators of the
filamentary structure of the Cosmic Web, the so-called \emph{global}
and \emph{local} skeletons. The first, based on a global study
of the matter density gradient flow, allows us to study the connectivity between a
density peak and its surroundings, with direct relevance to the
anisotropic accretion via cold flows on galactic halos. From the second,
based on a local constraint equation involving the derivatives of the
field, we can derive predictions for powerful statistics, such as the
differential length and the  relative saddle to extrema counts of the Cosmic web as a function of density
threshold (with application to percolation of structures and connectivity), as well as
a theoretical framework to study their cosmic evolution through the
onset of gravity-induced non-linearities. 
\end{abstract}

\maketitle

Over the course of the last decades,
our understanding of the extragalactic  universe 
has undergone a 
 paradigm shift:  the description of its structures  
 has evolved 
 from being (mostly) isolated to being multiply connected both on large scales, cluster scales and galactic scales.
  This interplay between large and small scales is driven in part 
 by the scale invariance of gravity which tends to couple dynamically different processes, 
 but also by a  the  strong theoretical prejudice associated with the so-called concordant cosmological model \citep{prunet}.
 This model  
 predicts a certain shape for the initial conditions, leading to a hierarchical formation scenario,
 which predicts the formation of the so-called Cosmic Web, the most striking feature of matter
distribution on megaparsecs scales in the Universe. This distribution
was confirmed, observationally, more than twenty
years ago by the first CfA catalog \citep{Lapparent}, and later by the
SDSS \citep{sdss} and 2dFGRS \citep{2df} catalogs. 
On these scales, 
the  ``Cosmic Web'' picture \cite{bkp}  relates
the observed clusters of galaxies, and filaments that link them, 
to the geometrical properties of the initial density field that are enhanced but not yet
destroyed by the still mildly non-linear evolution  \citep{zeldo70}. 
The analysis of the connectivity of this filamentary
structure is critical  to map the very large scale distribution of our universe 
 to establish, in particular, the percolation properties of the Web \citep{2000PhRvL..85.5515C}.

On intermediate scales,
the paradigm  shift is  sustained  by panchromatic observations of the environment of galaxies which illustrate sometimes  
spectacular merging processes, 
 following the pioneer work of e.g. \cite{Schweizer} (motivated by theoretical investigations such as  \cite{Toomre}).
The importance of anisotropic accretion on cluster and dark matter halo scales  
\citep{aubert,aubertpichon,Steinmetz}  is now believed to play a crucial role in regulating 
the shape and spectroscopic properties 
of galaxies.
Indeed it has 
been claimed (see  e.g. \cite{ocvirk, dekel}) 
that the geometry of the cosmic inflow on a galaxy (its mass, 
temperature and entropy distribution,  the connectivity of the local filaments network, etc.) is strongly correlated to its history and nature.
Specifically,  simulations suggest that cold  streams violently feed high redshift young galaxies with a vast amount of fresh gas, resulting in very efficient star formation.
One of the puzzles of galaxy formation involves understanding how galactic disks reform
after minor and intermediate mergers, a process which is undoubtedly controlled by anisotropic gas inflow.

Recently, 
\cite{SCP} presented
a method to compute
 the full hierarchy of the critical subsets of  a given
density field.  This approach connects the study of the
filamentary structure to the geometrical and topological aspects of the
theory of gradient flows \citep{jost}.
 In this paper, we focus on the connectivity of the corresponding
 network.
 Specifically,  since galaxy formation seems to be geometrically
 regulated by the accretion of cold gas from the LSS,
how can we use the skeleton to characterize this anisotropic accretion, and predict its evolution through perturbation theory ? 

\begin{figure}
 {\includegraphics[height=.3\textheight]{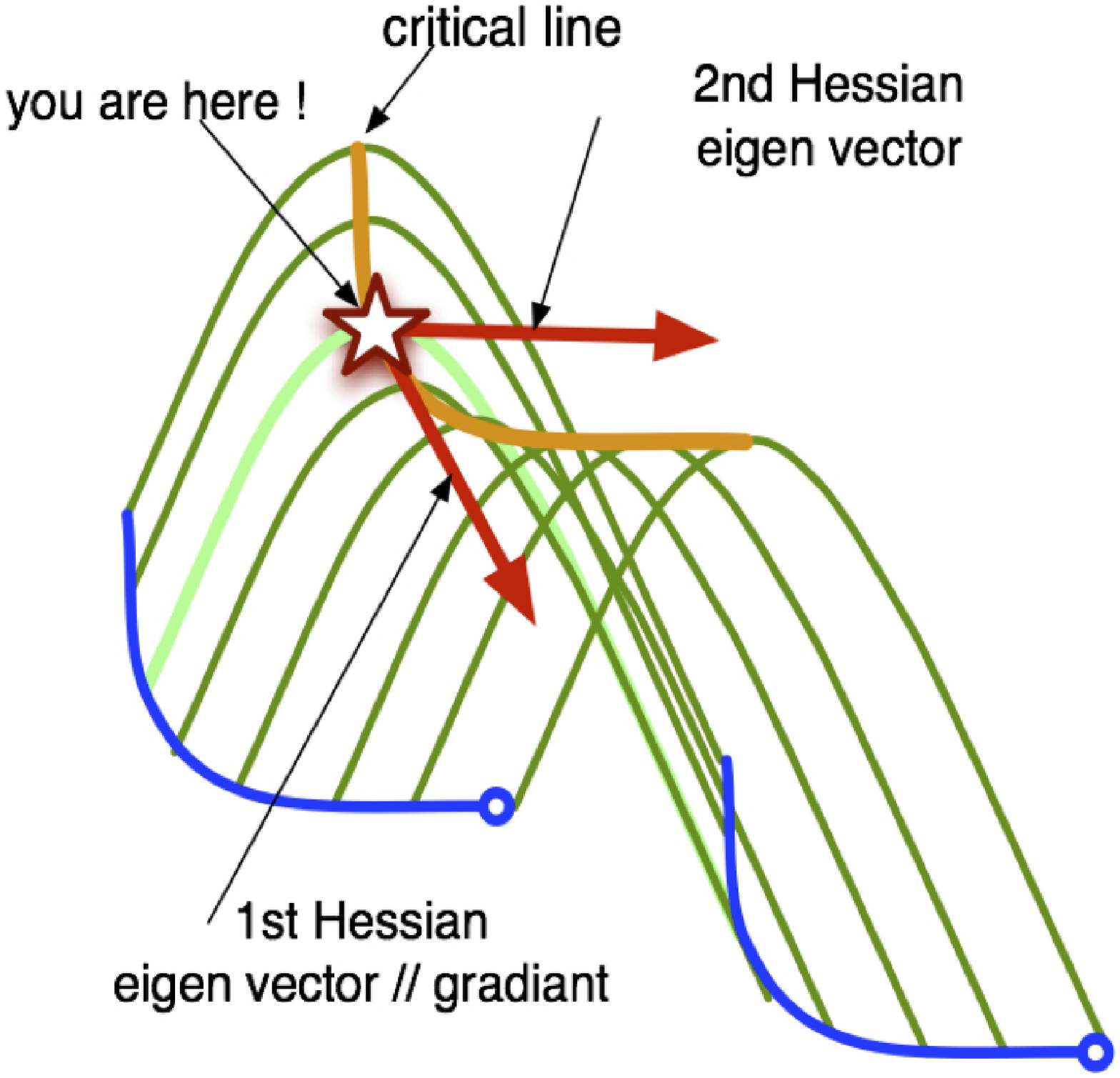}}
 {\includegraphics[height=.3\textheight]{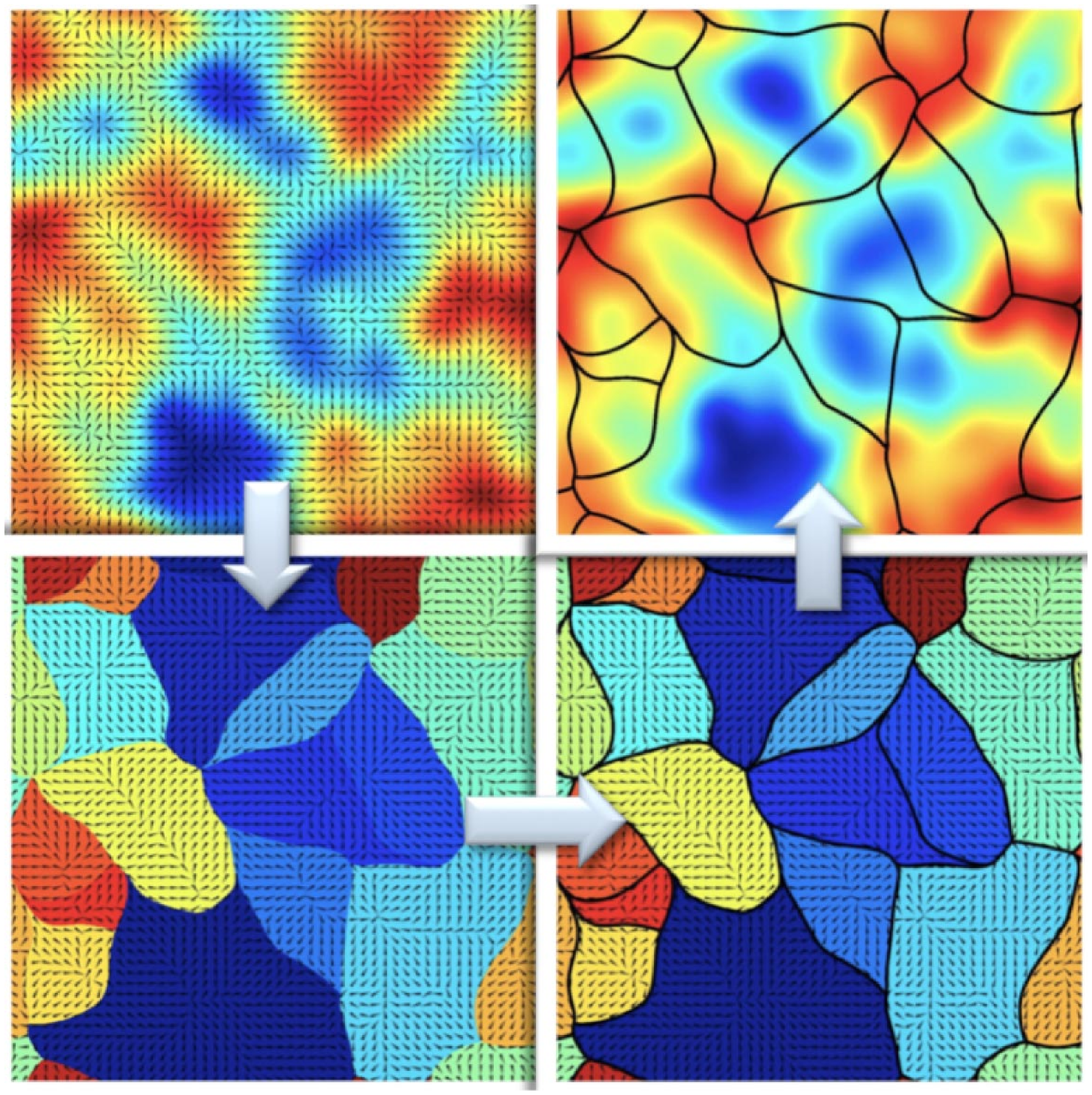}}
  \caption{
  {\sl Left:} an illustration of the local condition of the skeleton on a given ridge: on such a critical line, the gradient is parallel to the largest eigenvalue of the Hessian (corresponding to the least curvature).
  This defines a  degenerate point process which provides means of extending the critical point theory to critical lines.
   {\sl Right:}  the fully connected skeleton algorithm illustrated in 2D. Following the panels anti-clockwise, the gradient of a given field defines neighboring regions, which correspond to the set of pixels which lead to the same minimum (a so-called void-patch). The edge of these void-patches defines a set of lines which connect critical points together: the global skeleton. \label{fig:def}
  }
\end{figure}

Let us first qualitatively introduce the two operating skeleton extraction algorithms, and present our findings regarding the connectivity of random fields.
We will then summarize the underlying statistical theory, first for
Gaussian random fields, then for non-Gaussian fields, which allows us
to explain qualitatively
 the cosmic evolution of the connectivity of dark matter halos.
In doing so we will also demonstrate  how  the skeleton  applied to the large scale structure of the universe could be used to track $\Omega_{\rm DM}$
and  $D(z)$.

\begin{figure}
{\includegraphics[height=.35\textheight]{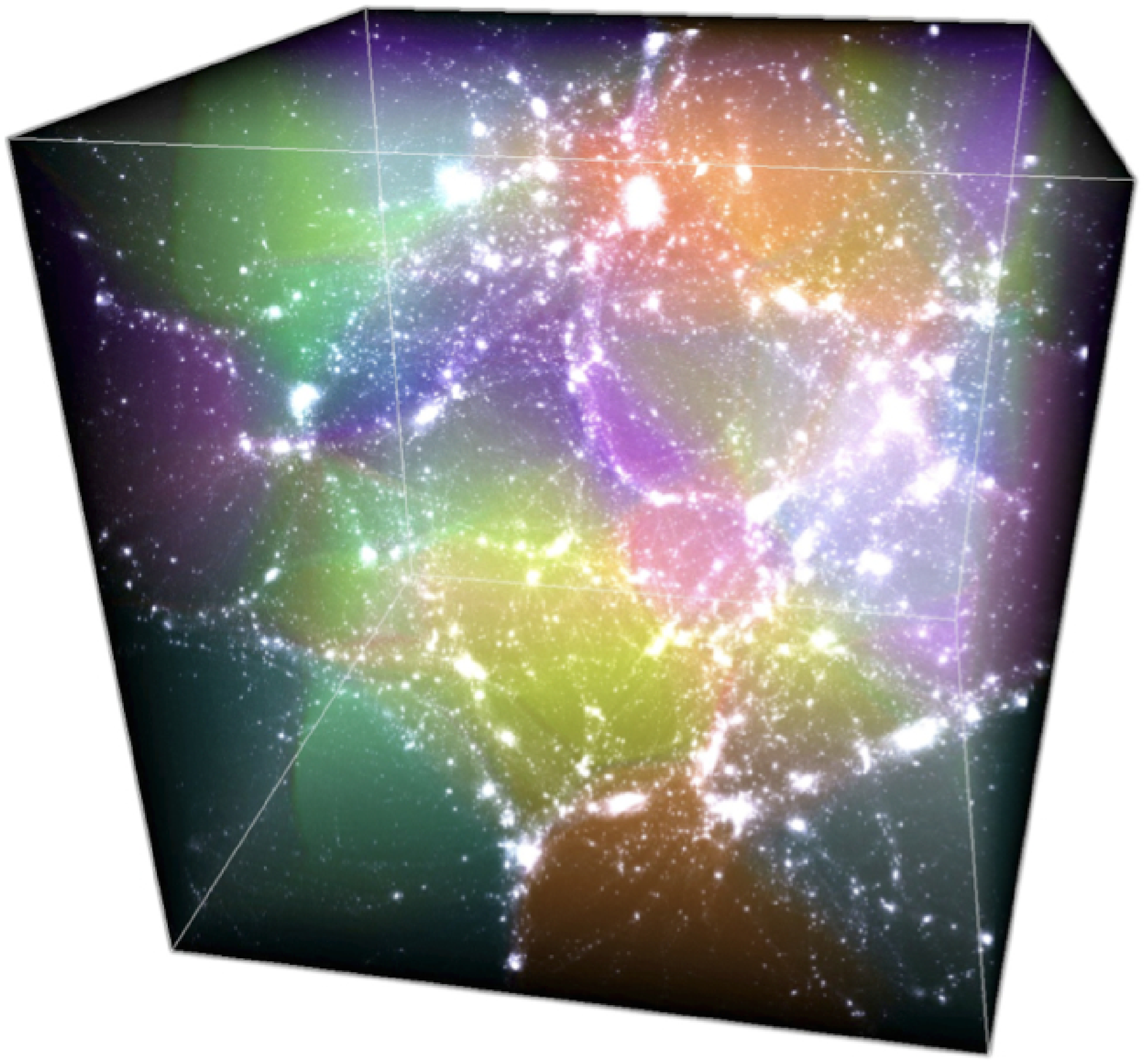}}
 {\includegraphics[height=.35\textheight]{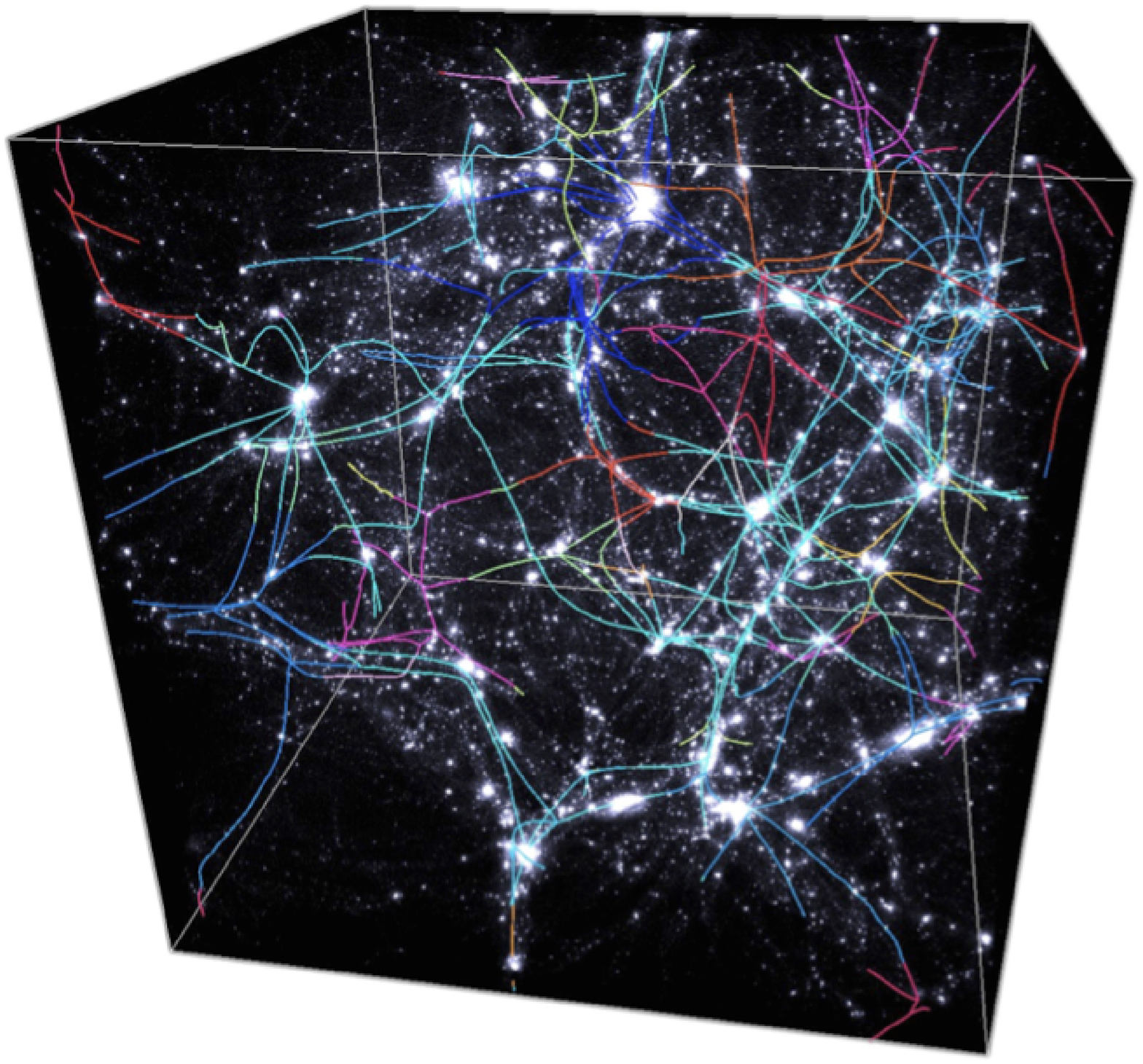}}
  \caption{
  {\sl Left: }  the 3D void-patch of a simulation of the cosmological density
density field in a $50h^{-1}$ Mpc box with the {\tt gadget-2} N-body code. 
This  void-patch segmentation was computed from a $128^3$ pixels sampling grid smoothed over $5$ pixels ($\approx 2 h^{-1}$ Mpc).
  {\sl Right: } the corresponding skeleton. The skeleton's different colours represent the index of the peak-patches (\emph{i.e.}, the void-patches of minus the density field) which provide by construction the natural segmentation of filaments attached to the different clusters.  \label{fig:def3D} }
\end{figure}

\subsection{The skeleton: algorithms}
\begin{figure}
 \includegraphics[height=.5\textheight]{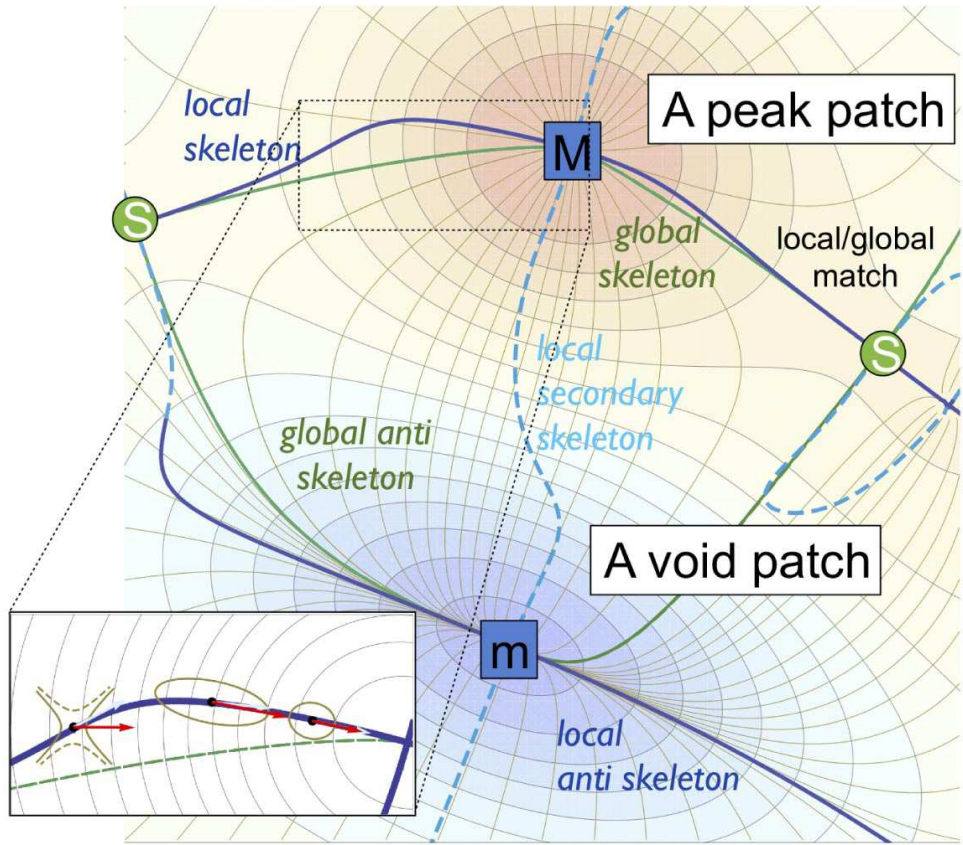}
 \caption{An example  of a generic patch of a 2D field.  
  The underlying isocontours correspond to the density field.
  The thin gold lines show the gradient lines of the field.
  The blue lines represent the local set of  critical lines, given by
  the solution of equation (\ref{eq:crit}). Primary lines are shown 
  in solid and secondary lines are dashed.
  The green lines correspond to global critical lines: the skeleton and 
  the anti-skeleton, which delineate a  special bundle of gradient
  lines at resp. the intersection of a peak-patch and a void-patch \citep{jost}.
  The primary local lines follow fairly well the gradient lines,
  noticeably near the saddle points, where the ``stiff'' approximation holds best.
  In contrast, the
  approximation worsens for the secondary critical lines.
  The main distinction between the global and local skeletons is that
  the global one follows everywhere the smooth gradient line that uniquely 
  connects a maximum to a saddle, at the cost of deviating from
  being exactly on the ridge (see how in the vicinity of the minimum at
  the bottom, the right green line does not follow the valley).
  The local skeleton tries to delineate the ridges as far from extrema
  as possible, but then the lines that follow this local
  procedure from different extrema do not meet and have to rather suddenly
  reconnect. 
  A particularly striking example is the loop on the right hand side.
 {\sl Bottom left medallion:} the neighbourhood of a local critical line (thick blue line).
  Thin lines are isocontours of the field. Three sample points are
  investigated in detail. The signature, orientation
  and the magnitude of the local Hessian are represented
  by the golden shapes. Near the maximum on the right edge, the signature
  of the eigenvalues of the Hessian is (-1,-1), which is shown by ellipses
  oriented according to eigen-directions with longer semi-axis along the 
  direction of the least curvature. At the leftmost point the eigenvalue 
  signature is ``saddle-like'', (1,-1), which is represented by a pair of 
  hyperbolae, also oriented with respect to eigen-directions.
  By definition, on the critical line the gradient of the field $\nabla \rho$, 
  shown by red arrows, is aligned with one of the eigen-directions (i.e the
  axis of the ellipse or hyperbola in the graph).
  The light cyan arrows are the tangent vectors to the critical line
  $ \propto \boldsymbol{\epsilon} \cdot \nabla S$, while stiff approximation
  to them would be parallel to the gradient. The direction of the critical line
  is close to the gradient when it follows the ridge near the maximum,
  but slides at an angle in the ``saddle-like'' region, before joining the
 corresponding  saddle extremal point. Note that the gradient line
  that takes us to the same saddle as a segment of the global skeleton
  (dashed green) does not follow the ridge too closely in this instance.
   \label{fig:local-global}
}
\end{figure}

We may qualitatively define the skeleton as the 3D analog of ridges in a mountainous landscape.
It consists of the special lines which connect saddles and peaks together. 
Mathematically these are critical  solutions of the gradient flow ($\dot r= \nabla \rho$) between critical points. 
Here we want to construct such 3D ridges to trace the filaments of the large scale structures.
Two venues have been explored recently: the so called  local skeleton \cite{skel2D,skllocal}, which defines a (degenerate) point-like process corresponding to the zeros of a set of functions; 
the idea is that on a ridge the gradient should be along the direction of least curvature (see figure~\ref{fig:def}, left panel), which translates into the set of 3 equations:
\begin{equation}
\mathbf{S} \equiv \left( \nabla \nabla \rho \cdot \nabla \rho \right) \times \nabla \rho = \mathbf{0}\,, \label{eq:crit}
\end{equation}
where $\nabla \rho$  and $\nabla \nabla \rho$ are respectively the gradient and  the Hessian of the field, $\rho$ and $\lambda_1>\lambda_2>\lambda_3$ its eigenvalues,
with the condition $\lambda_1+\lambda_2 \leq{0}$, which ensures that the critical line is a ridge. Alternative conditions on the eigenvalues of the Hessian may be applied to
pick up the whole set of critical lines, see figure~\ref{fig:local-global}.   Following \cite{skel2D}, this local skeleton and its properties provide an
alternative description 
to  classical approaches of galaxy
clustering (powerspectrum, bispectrum {\sl etc.})  and attempt
to achieve data compression via a 
mathematical description of the morphology of the cosmic web.

{An alternative, global definition \cite{skel2D,SCP} is to define it as the border between $N$ void-patches, where N is the dimensionality and void-patches are the attraction basins of the (anti-)gradient of the field (see figure~\ref{fig:def}, right panel in 2D, and figure~\ref{fig:def3D} in 3D). 
The actual  implemented algorithm  is based on a watershed technique and uses a
probability propagation scheme to improve the quality of the
segmentation (see also \cite{Rien1}). 
It can be applied within spaces of arbitrary dimensions and geometry.
The corresponding
recursive segmentation yields the network of  the primary critical
lines of the field: the fully connected skeleton that continuously
link maxima and saddle-points of a scalar field together.
Both constructions are of interest: the local formulation  provides means of conducting detailed statistical
investigation of the corresponding  degenerate point-like process, while the global skeleton
 yields a totally connected network of lines.
 }
This critical set of lines is a compact description of the geometry of the field, richer than the knowledge of the critical points alone.
For the purpose of this review, it also allows us to explore the connectivity of peaks, both from the point of view of the number of connections to other peaks, and  of the 
number of skeleton branches leaving a given maximum. Conversely, the local skeleton formulation allows us to extend the BBKS theory \cite{BBKS} of peaks in the context of critical lines to investigate their cosmic evolution, see below.
Figure~\ref{fig:local-global} compares the two sets of lines in the neighbourhood of a given peak.

\subsection{The skeleton: connectivity}
\begin{figure}
 {\includegraphics[height=.65\textheight]{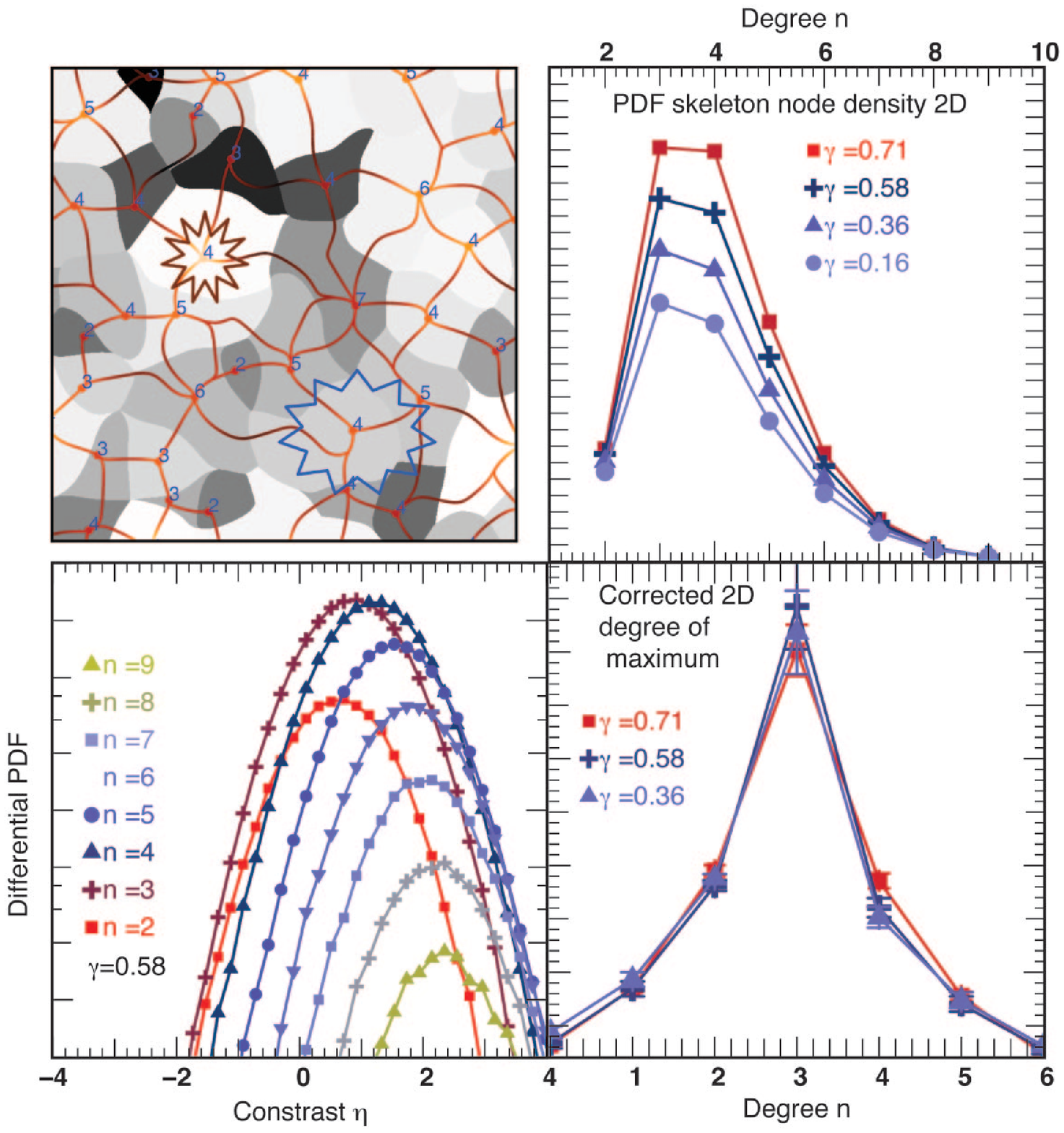}}
  \caption{ { \sl Top left panel}: A typical (gray coded) 2D peak-patch  of a Gaussian random field together with the degree of each vertex of the skeleton graph; for instance, within the red star
  the mean degree is 4, while the corrected degree is 3 within the blue star, as one bifurcation point is present;
 { \sl top right panel:} the corresponding PDF of this degree, for different values of  the shape parameter $\gamma$ as labeled; the mean degree is 4 and corresponds to
the theoretical expectation of $2n_{\rm saddle}/n_{\rm max}$;
{ \sl bottom left panel:}  the (log) differential PDF of the degree as a function of contrast: as expected, the denser peaks are more connected; interestingly, the distribution
seems Gaussian at fixed $n$ even though its marginal isn't;
 { \sl bottom right panel:}  the corrected PDF of the number of branches connecting onto a given peak, which involves substracting the number of bifurcation points within
 the patch to the number of saddle points on the edge of the patch. \label{fig:connectix2D}
 }
\end{figure}

Let us now make use of this algorithm to explore the connectivity of the corresponding network.
A set of  two dimensionnal Gaussian random fields (GRF) is produced, 
and for each of them, the peak patch and  the skeleton was generated and its connectivity computed, following 
the second prescription described in \cite{SCP}: we chose here to smooth and label the skeleton having
fixed  the field extrema, (i.e. {\sl not}  the  bifurcation points of  the
skeleton, see Figure~\ref{fig:connectix2D}  ({\sl top left panel})).  By fixing the extrema of the field, one
ensures that the skeleton subsets that link these extrema are treated
independently. 

Let us first focus on the degree of the peakpatch, defined as the number of saddle points within a given patch which connects 
the skeleton of one patch to its neighbours.
Hence the  connectivity count  reflects the number of maxima a given peak is connected 
to (or in the language of graph theory, the most likely degree of the vertices: $\hat n =\langle n \rangle$).
Figure~\ref{fig:connectix2D}  ({\sl top right panel}) displays this PDF as a function of $n$; in particular it allow us to compute $\langle n \rangle$,
the mean $n$ which is found to be independent of $\gamma$, the shape parameter of the underlying GRF.
The {\sl  bottom left panel} shows the corresponding  normalized PDF$(n,\eta)$ versus the contrast $\eta$.
As expected the number of connections increases with contrast. 
Indeed, near the maximum at high contrast all eigen values tend to become equal \citep{PB}. Therefore all incoming
directions become possible.

\begin{figure}
 {\includegraphics[height=.35\textheight]{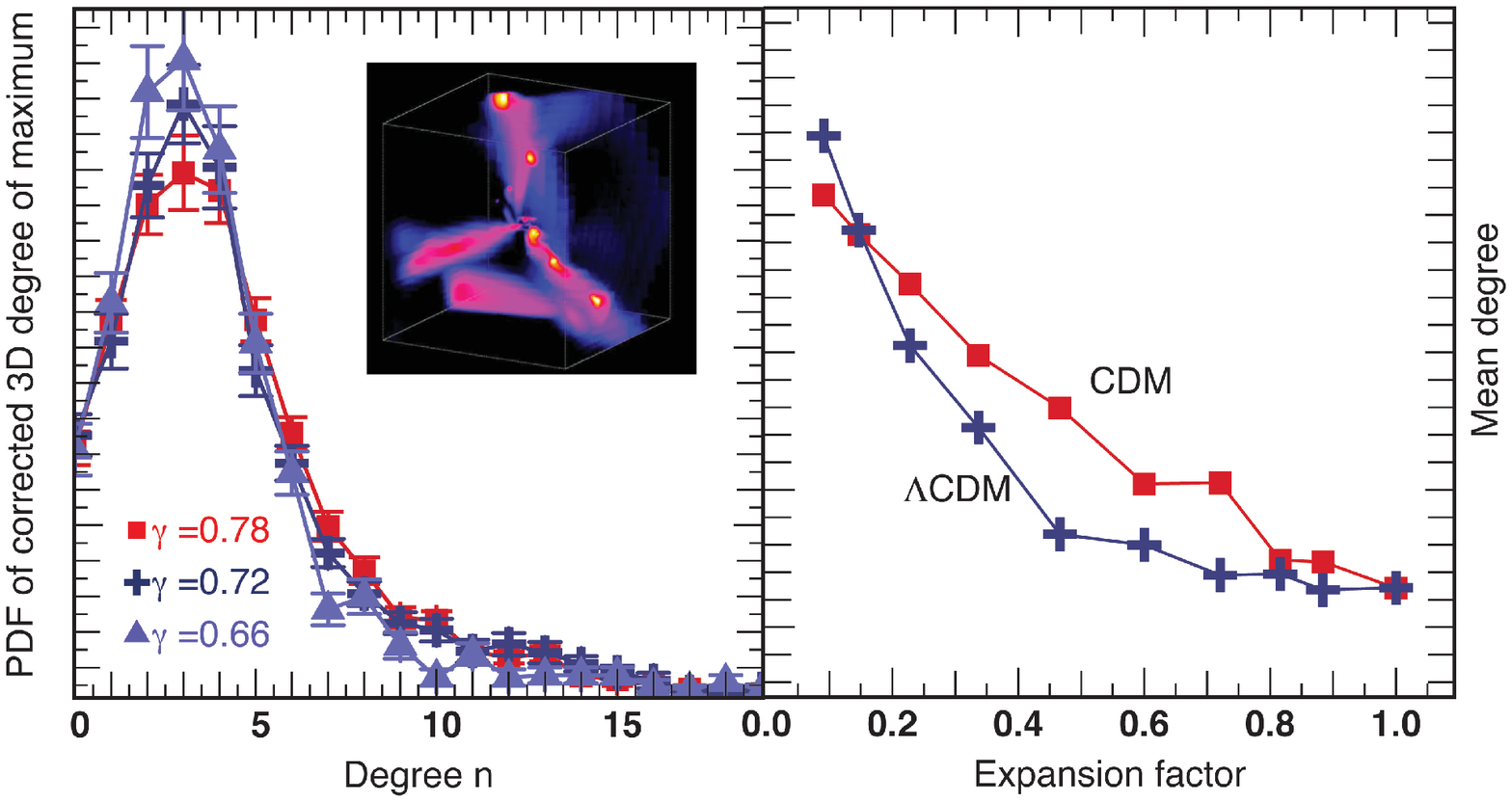}}
  \caption{ { \sl Left panel}: the 3D corrected PDF of the number of branches connecting onto a given peak for a GRF: the mode is at 3; {\sl insert}: 
  the typical neighbourhood entropy map of a massive galaxy in the Marenostrum simulation from \cite{dekel}; 
 { \sl right panel:}   the  qualitative cosmic evolution of the mean degree in dark matter simulation with and without dark energy; as expected this degree decreases with 
 the expansion factor as connectors get washed out by gravitationnal clustering;
the rate of decrease seems sensitive to the moment when dark energy  kicks in for $\Lambda$CDM models (around $a\sim 0.5$). Note that the absolute amplitude of this mean
degree is not shown here as we expect the global skeleton to over estimate the mean connectivity in 3D (see main text).  \label{fig:connectix3D}
}
\end{figure}

Let us now take the other (astrophysically relevant) perspective and count the actual number of filaments incident onto a given 
maximum. This local (intra patch) degree is in fact equal to the number of saddle points minus the number of bifurcation 
within the peakpatch:
$
n_{\rm max}=n_{\rm saddle} -n_{\rm bifurcation}\,.
$
The {\sl  bottom right panel} of figure~\ref{fig:connectix2D} shows the corresponding PDF($n_{\rm max}$). Note that this distribution is almost symmetrical, centered  at 
$\langle n_{\rm max}\rangle=3$.
Figure~\ref{fig:connectix3D}  ({\sl left panel}) presents the same distribution in 3D, which is very skewed and presents a sharp mode at 3.
Its cosmic evolution, measured in CDM dark matter simulations with and without dark energy is qualitatively shown on the   {\sl right panel}.
Our purpose  in the rest of this  paper is to explain qualitatively this cosmic trend by 1) deriving  the statistics of bifurcation points  within each patch and 2) predicting 
 the cosmic evolution of saddle points and peaks. 

\subsection{The skeleton: statistics}

\begin{figure}
 {\includegraphics[height=.275\textheight]{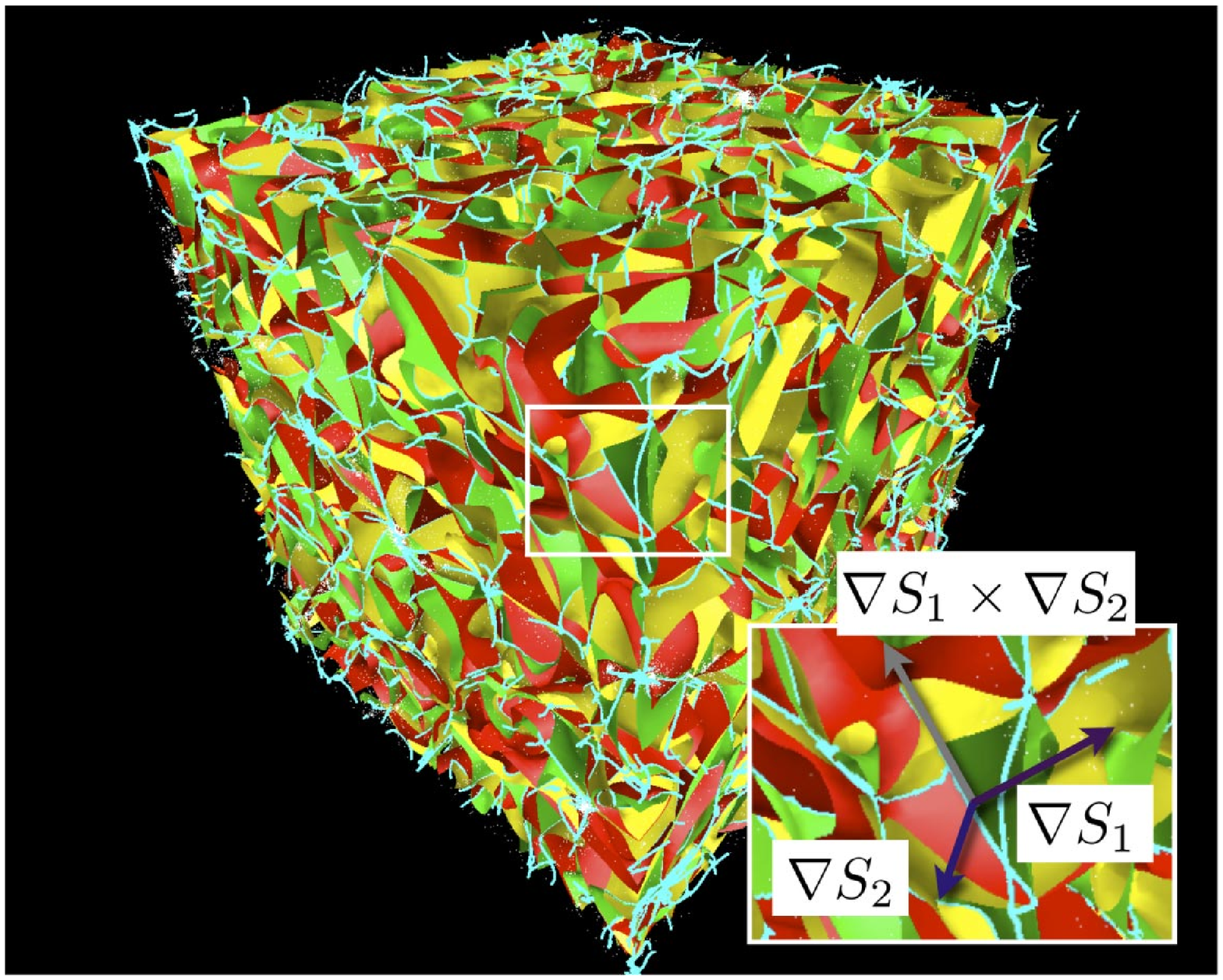}}
 {\includegraphics[height=.275\textheight]{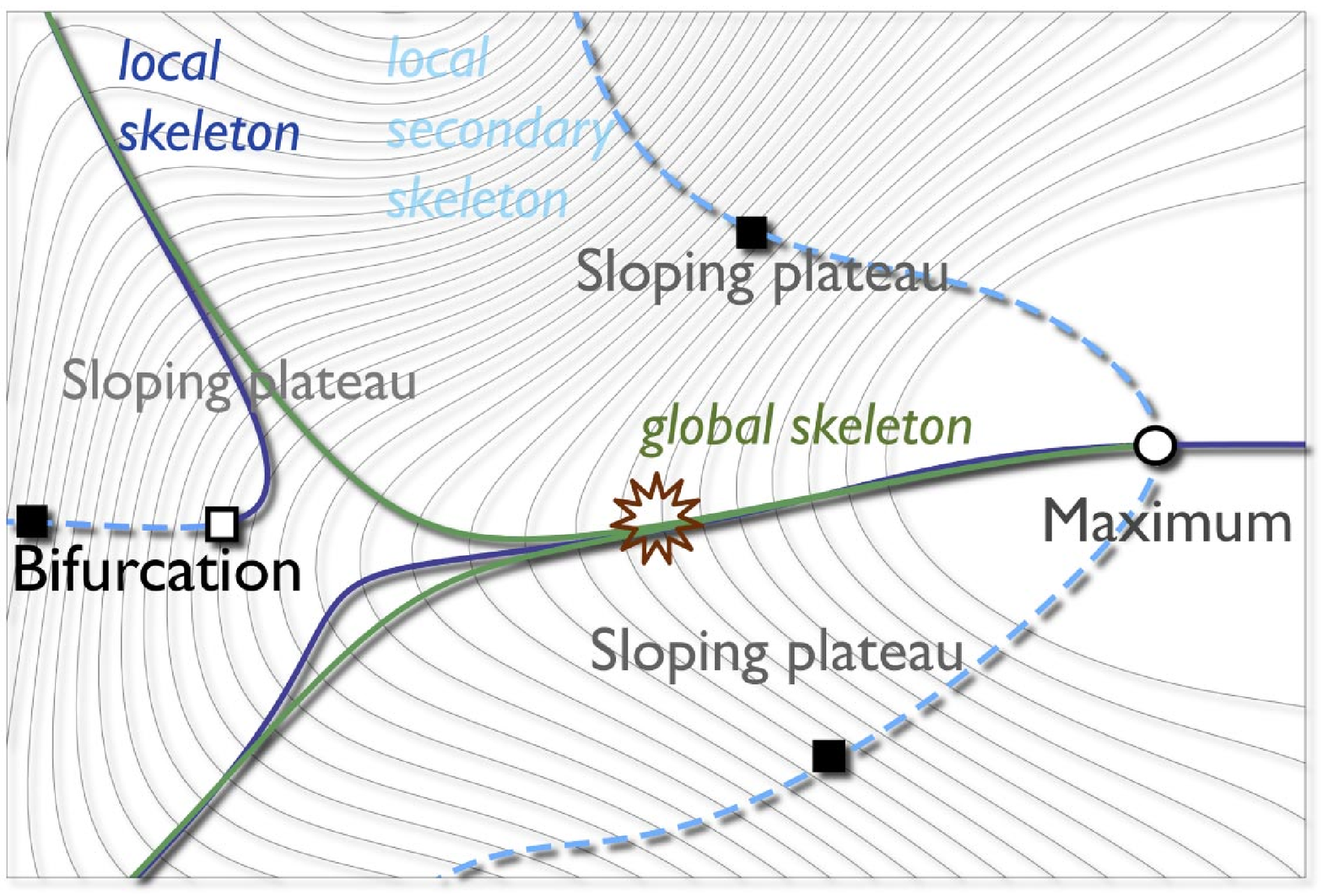}}
  \caption{{\sl left panel}: 
  the local direction of the skeleton is along the cross product, $\nabla S_1 \times \nabla S_2 $ of the normals, $\nabla S_1$ and $
  \nabla S_2$, of the iso-surfaces, $S_1$ and $S_2$, defined by the critical condition~(\ref{eq:crit}). Hence the differential length per unit volume of the skeleton
   is given by
  expectation of the modulus of this cross product, which can be expressed locally with the JPDF of the field and its successive derivatives. \label{fig:skeldir}
  {\sl right panel}: the three types of singular point ($ \nabla_k S=\mathbf{0}$) on the critical
   lines (in solid blue: primary; dashed: secondary; green: gradient
   lines of the global skeleton):
      an extremum ({\sl open circle}), a ``bifurcation'' ({\sl white square}),
      and ``slopping plateaux'' ({\sl black squares}). The red star corresponds to what would naively correspond to the bifurcation point.
      \label{fig:bif}
}
\end{figure}

As illustrated first in 3D on Figure~\ref{fig:skeldir},  the critical condition given by equation~(\ref{eq:crit}) defines 3 isosurfaces corresponding to its components, 
and the local skeleton direction is simply given by the cross products of two of its normals. 
Hence the {\it differential} length (per unit volume) is simply given by the statistical expectation
\begin{equation}
\frac{\partial{\cal L}}{\partial \eta}=\left( \frac{1}{R_*}\right)^2
\int  d^{3}x_{k} d^{6} x_{kl}  d^{10} x_{klm} \
|\nabla s^i \times \nabla s^j| {P}(\eta,x_{k},x_{kl},x_{klm})
\delta_{D}\left(s^i(x_{k},x_{kl},x_{klm})\right)
\delta_{D}\left(s^j(x_{k},x_{kl},x_{klm})\right) \,,
 \label{eq:myeq2}
\end{equation}
where ergodicity  allowed us to replace volume average by ensemble average over the statistical distribution, ${P}(\eta,x_{k},x_{kl},x_{klm})$, of the successive (a-dimensional) derivatives, 
 $\sigma_1 x_{k}\equiv \nabla_k \rho$,
$ \sigma_2 x_{kl}\equiv \nabla_k \nabla_l \rho,$ $
\sigma_3 x_{klm}\equiv \nabla_m \nabla_l \nabla_k \rho$ of the field, $\rho$.
 The scale factor ${1}/{R_*}^2= \sigma_2^2/\sigma_1^2$ in equation~(\ref{eq:myeq2}) follows dimensionally (given $ \sigma_2^2 \sigma_1 \nabla s^i \equiv \nabla {S}^i$, $\sigma_0 \eta \equiv \rho$). 
Here the variances, $ \sigma_p^2$,  obey $  \sigma_p^2 ={2 \pi^{N/2}}/{\Gamma[N/2]}
  \int_0^\infty k^{2p} P(k) k^{N-1} dk$  in $N$ dimensions. (see \cite{skllocal} for a formal derivation of equation~(\ref{eq:myeq2})). 

More generally, for ND critical lines, the N-1 independent functions 
$S^i$ that define the critical condition (\ref{eq:crit}) acquire
the following  reduced form in the eigenframe of the Hessian of the field:
$
s^i= x_1 x_i \left(\lambda_1-\lambda_i\right)=0.
$
The measurements in \cite{SPCNP}
found  that, over the range of
spectral indexes relevant to cosmology, the third derivatives of the field $x_{klm}$ play a subdominant 
nature. In the so called  ``stiff'' approximation
we therefore omit the third derivative, 
effectively assuming that the Hessian can be treated
as constant during the evaluation of $\nabla s$.
This picture corresponds to a  skeleton connecting extrema with relatively straight segments. Let us first assume that the underlying field is Gaussian.
In the stiff approximation, the gradients ${s^i}_k \equiv
\nabla_{} s^i $ have just two non-zero components,
${s^i}_1= x_i \lambda_1 (\lambda_1-\lambda_i)$
(which vanishes on the critical line) 
and ${s^i}_i = x_1 \lambda_i (\lambda_1-\lambda_i) $, which shows that in this approximation we equate the direction of the
line with the gradient of the field. Substituting this expression into
equation~(\ref{eq:myeq2}) and integrating over $\delta_{\rm D}(s^i)=
 \delta_{\rm D}(x_i)/({x_1 (\lambda_1-\lambda_i)})$
we obtain a simple expression for the differential length of the ND-critical lines: 
\begin{equation}
\frac{\partial {\cal L}^{\mathrm{ND}}}{\partial \eta}\propto
\left( \frac{1}{R_{*}}\right)^{N-1}\frac{1}{\sqrt{1-\gamma^2}}
\int \cdots \int \prod_{i\le N}  d \lambda_i 
 \prod_{i<j} (\lambda_i-\lambda_j)
\left| \prod_{i>1}   \lambda_i \right|  \exp\left( -\frac{1}{2}
Q_\gamma(\eta,\{\lambda_{i}\})
\right)\,, \label{eq:NDdiff}
\end{equation}
where the shape parameter $\gamma  ={\sigma_1^2}/
{(\sigma_0\sigma_2)} $ describes the correlation between the field and its second derivatives,  $Q_{\gamma}$ is a quadratic form in $\lambda_{i}$ and $\eta$ which
functional form is
\begin{equation}
Q_{\gamma}(\eta,\{\lambda_{i}\})=\eta^2+ \frac{\left(\sum_{i}\lambda_{i}+\gamma \eta\right)^2}{ (1-\gamma^2)}+
 (N+2)\left[\frac{1}{2}(N-1) \sum_{i} \lambda^{2}_{i}- \sum_{i\neq j} \lambda_{i}\lambda_{j}\right].
\end{equation}

\begin{figure}
 {\includegraphics[height=.25\textheight]{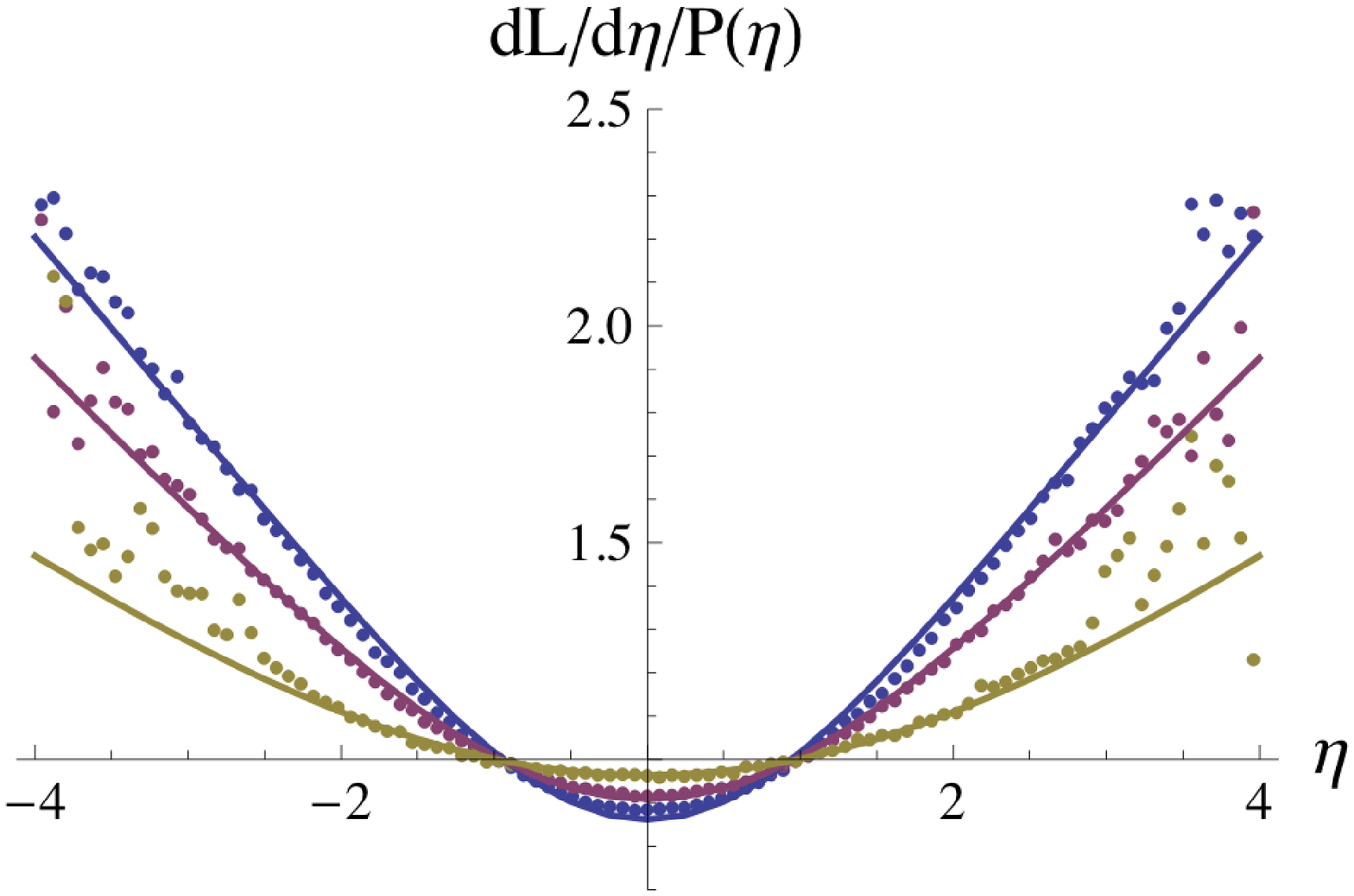}}
 {\includegraphics[height=.25\textheight]{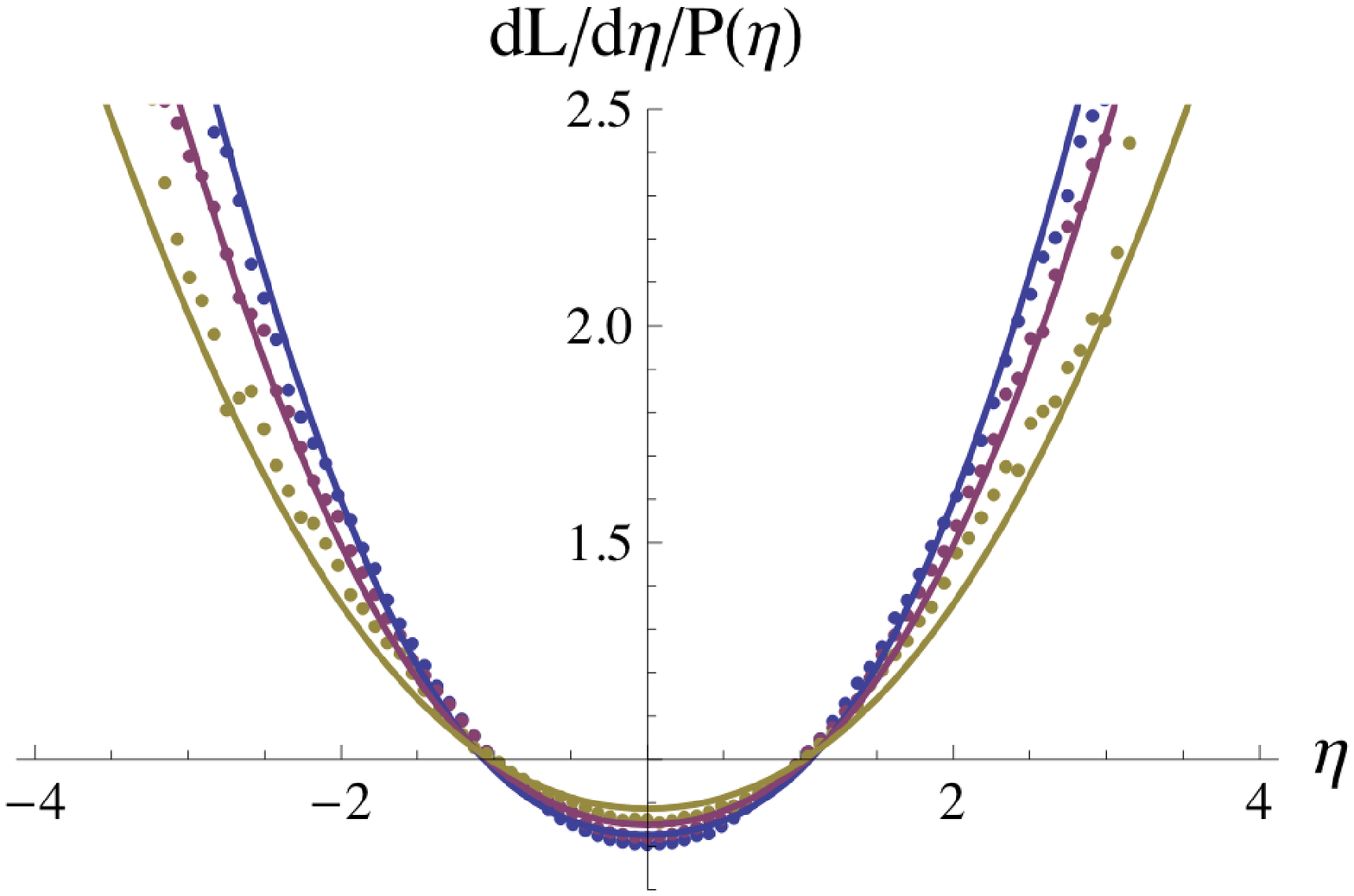}}
  \caption{The relative differential length,
${\partial {\cal L}}/{\partial \eta}/\mathrm{PDF}$,
measured in simulation of 2D ({\sl left}) and 3D ({\sl right}) Gaussian
random fields with scale invariant power-law spectra versus predictions of
the local theory in stiff approximation (solid curves).
The spectral parameters are $\gamma= 0.71, 0.59, 0.39$ for the 2D and
$\gamma=0.77,0.70,0.60$ for the 3D simulations.}
\label{fig:DiffLen}
\end{figure}

Equation~(\ref{eq:NDdiff}) simply states that the stiff differential length is the expectation of the product of the ``smaller" eigenvalues, which is quite reminiscent of 
the classical extremal count result (the expectation of the product of  all eigenvalues). It also qualitatively makes sense, as the larger the curvarture orthogonal to the 
skeleton, the more skeleton segment one may pack per unit volume.
Since the argument of $Q_\gamma$ is extremal as a function of $\eta$ when
$\gamma \eta \sim \sum_i \lambda_i$, the largest contribution at large
$\gamma \eta$ in the integral should arise when $\lambda_i \propto \gamma \eta$
since near the maximum at high contrast all eigen values are equal \citep{PB}. 
Hence given that $\prod_{i<j} (\lambda_i-\lambda_j)$ is  the measure, the only
remaining contribution in the integrand comes from
$\left| \prod_{i>1}   \lambda_i \right|  \propto (\lambda \eta)^{N-1}$, and
the dominant term at large $\eta$ is given by 
\begin{equation}
\frac{\partial {\cal L}^{\mathrm{ND}}}{\partial \eta}\stackrel{\gamma \eta \to \infty}{\sim}
\frac{1}{\sqrt{2 \pi}} \exp \left[-\frac{1}{2} \eta^ 2\right] \left(\frac{\eta}{R_0}\right)^{N-1}\,,\quad {\rm with } \quad R_0=\frac{\sigma_{0}}{\sigma_{1}}\,.
\label{eq:asymptotic}
\end{equation}
Note that in 2D  the differential length, equation~(\ref{eq:NDdiff}) can be reduced to  a  particularly simple closed form
\begin{equation}
\frac{\partial {\cal L}^\mathrm{skel}}{\partial \eta} =
\frac{1}{\sqrt{2 \pi}} \exp\left[-\frac{\eta^2}{2}\right]
\left[\frac{1}{8} \left( 1+\frac{2}{\sqrt{\pi}} \gamma \eta \right) 
\left(1 + \mathrm{Erf}\left[\frac{\gamma \eta}{\sqrt{2} \sqrt{1-\gamma^2}}\right] \right)
+\frac{\sqrt{1-\gamma^2}}{2 \sqrt{2} \pi} 
\exp\left(-\frac{\gamma^2 \eta^2}{2 (1-\gamma^2)}\right)
\right]\,,
\label{eq:skeleton}
\end{equation}
and for the integrated skeleton length
\begin{equation} 
L^\mathrm{skel} = \frac{1}{8}+\frac{\sqrt{2}}{4 \pi}
 = 0.23754 ~ (\times R_*^{-1}) \quad .
\label{eq:skeleton_tot}
\end{equation} 
 {In other words, one expects
to find one segment of skeleton per linear section of $\approx (4.2 R_*)$.} Similarly, in 3D one segment of skeleton is found per surface section of  $\approx (4.65 R_*^2)$.
The match of the detailed PDF with the corresponding measurements are shown in Figure~\ref{fig:DiffLen} in 2 and 3D.
From the point of view of cosmology, the skeleton is invariant w.r.t. any monotonic bias,  and traces the denser regions of the field.
As shown in equations~(\ref{eq:asymptotic}) and (\ref{eq:skeleton_tot}) its statistical description yields a measure of both $R_*$ and $R_0$, hence on the shape of the underlying 
powerspectrum on  the corresponding smoothing scale over which the skeleton was computed. An implementation of this estimate on the SDSS catalog was carried by \cite{skellet}
and provided constraints on $\Omega_{\rm M}$.

\subsubsection{Bifurcation counts}
In \cite{pogoskel} we conjectured that critical lines experience a qualitative change in 
behaviour in the vicinity of the points where either a Hessian eigenvalue
orthogonal to the gradient direction vanish, or becomes 
equal to the one along the gradient. 
The first type corresponds to points where the curvature transverse to
the direction of the critical line vanishes along at least one axis:
typically, in 2D, they mark regions where a crest becomes a trough, or 
vanish into a plateau.
The second type  correspond to  points where the critical lines would split,
even though the field does not go through an extremum: a bifurcation
of the lines occurs along the slope; the occasional skier or mountaineer will
be familiar with a crest line splitting in two,  even though the gradient of
the field has not vanished. 
Namely, if, for definiteness, 
$\nabla \rho$ is taken to be along the first eigen-direction, $\lambda_2=0$, or 
$\lambda_2=\lambda_1$.  
We called the first case the ``sloping plateau'' as it 
designates the entering of a flat region, and the second, 
the ``bifurcation'' as it designates the places of possible reconnection
of critical lines.
Remarkably, these special points on the critical lines
are recovered by the formal singular condition zero-tangent vector
if $\nabla_k S^i$ is evaluated in the stiff approximation. Along the ND critical line defined by
$x_2=\ldots=x_N=0$, the nul tangent vector condition, $
x_1^{N-1} \prod_{i>1} \lambda_i \left(\lambda_1-\lambda_i\right) = 0 $
gives rise to three classes of situations: (i) $x_1=0$ corresponding to
extremal points; (ii) one of $\lambda_i=0$ corresponding to
slopping flattened tubes; and (iii) one of $\lambda_i=\lambda_1$,
corresponding to an isotropic bifurcation.

To be more specific, in 2D, the skeleton's singular points correspond to points where 
$S_k \equiv \nabla_k S=\mathbf{0}$.
The number density, $n_{{\rm B}}(\eta)$ of singular points below the
threshold $\eta$ is equal to
\begin{equation}
n_{{\rm B}}(\eta) = \int_{\eta>x}   d x d^{2}x_{k} d^{3} x_{kl}  d^{4} x_{klm}  d^{5} x_{klmn} P(x,x_{k},x_{kl},\cdots)
|{\rm det}\left(\nabla_k \nabla_l s\right)|\delta_{{\rm D}}(s_1)
\delta_{{\rm D}}(s_2)\,. 
\end{equation}
The gradient of $S$, evaluated in the stiff approximation,
in the Hessian eigenframe has the components 
$
s_1^\mathrm{stiff} = x_{2} \lambda_1 \left(\lambda_1-\lambda_2\right),
$ {\rm and} $
s_2^\mathrm{stiff} = x_{1} \lambda_2 \left(\lambda_1-\lambda_2\right),
$
and involves only second derivatives of the field. 
Let us consider the critical line that corresponds to the
$x_2=0$ condition in the Hessian eigenframe. Then $s_1^{\mathrm{stiff}}$
vanishes everywhere along this line. The requirement $s_2^{\mathrm{stiff}}=0$
has a solution at the extremal points, $x_1=0$, but also in
two other cases, namely $\lambda_2=0$ or $\lambda_2=\lambda_1$, that
we conjectured to be of interest.
The second situation (isotropic Hessian) has a number density given by  
\begin{eqnarray}
\frac{\partial n_{\rm B}^{\rm I}}{\partial \eta}= \frac{1}{ {\pi \tilde R}^2} 
\left[\frac{1}{\sqrt{2 \pi}} \exp\left(-\frac{\eta ^2}{2}\right)\right]
\left[ \frac{2}{\sqrt{2-\tilde\gamma^2}} -\frac{1}{2} (1+ {\tilde \gamma}^2) 
\right]\,,\quad{\rm with} \quad  \tilde R  = \frac{\sigma_{2}}{\sigma_{3}} \quad {\rm and}\quad {\tilde \gamma}=\frac{\sigma^2_2}
{\sigma_3 \sigma_1}\,.
\label{eq:nB2D-2}
\end{eqnarray}
We note that the number density of ``bifurcation'' points is proportional just 
to the PDF of the field and, consequently the bifurcation points
are as frequent in the regions
of high field values as in the low ones.   
Finally, in 3D,  we expect  the finite resolution bifurcation branch (connecting maxima to bifurcation points) to become bifurcation plates; in turn 
the boundary of these plates will be counted as critical lines by our voidpatch algorithm; we therefore expect that the algorithm numerically over estimate the 
degree of peakpatches.  Note nonetheless that the 3D counterpart of equation~(\ref{eq:nB2D-2}) should allow us to correct for the number of bifurcations within each patch
and compute the statistics of the number of branches connected to a given maximum.

\subsubsection{Departure from Gaussianity}
While the Gaussian limit provides the fundamental starting point in the study
of random fields \cite{Adler,Doroshkevich,BBKS},
non-Gaussian features of the CMB and the large scale structure (LSS) fields are also of great interest.
CMB inherits high level of Gaussianity from the
initial fluctuations, and small non-Gaussian deviations may
provide a unique window into the details of processes in the early Universe.
The gravitational instability that nonlinearly maps the initial Gaussian 
inhomogeneities in matter density into the LSS, on the other hand,
induces strong non-Gaussian features culminating in the formation of
collapsed, self-gravitating objects such as galaxies and clusters of galaxies.
At supercluster scales where non-linearity is mild, the non-Gaussianity of
the matter density is also mild, but still essential for quantitative 
understanding of the filamentary Cosmic Web \cite{bkp}
in-between the galaxy clusters.

In order to extend the result of the previous section, following \cite{PGP},
let  us develop the equivalent of the Edgeworth expansion
for the JPDF of the field variables that are invariant
under coordinate rotation. Such distribution can be obtained directly from general principles:
{ the moment expansion of the non-Gaussian JPDF corresponds
to the expansion in the set of polynomials which are orthogonal with respect
to the weight provided by the JPDF in
the Gaussian limit}. Thus, the problem is reduced to finding such polynomials
for a suitable set of invariant variables.
The rotational invariants that are present in the problem 
are: the field value $\eta$ itself, the modulus of its gradient, 
$q^2=\sum_{i} {x_i}^2$ and the invariants of the matrix
of the second derivatives $x_{ij}$.
A rank N symmetric matrix has N invariants with respect to rotations.
The eigenvalues $\lambda_i$ provide one such
representation of invariants, however
they are complex algebraic functions of the matrix components. An alternative, more useful 
representation is given by the linear combination of the polynomial invariant,  
$I_s$ (where the linear invariant 
is the trace, $I_1=\sum_{i} \lambda_i$, 
the quadratic one is $I_2=\sum_{i<j} \lambda_i \lambda_j$
and the N-th order invariant is the determinant of the matrix.
$I_N=\prod_{i} \lambda_i $)
$J_1 = I_1 ~, 
J_{s\ge 2}  = I_1^s - \sum_{p=2}^s{(-N)^p C_{s}^p}/{(s-1)/C_N^p} I_1^{s-p} I_{p}\,,
\label{eq:Js}
$
where $J_{s\ge2}$ are (renormalized) coefficients of the characteristic
equation of the {\it traceless part} of the Hessian and are independent
in the Gaussian limit on the trace $J_1$. 
Let us consider again here  the 2D  case explicitly.
Introducing $\zeta=(
\eta + \gamma J_1)/\sqrt{1-\gamma^2}$ in place of the
field value $\eta$ we find that the 2D Gaussian JPDF 
$G_{\rm 2D}(\zeta ,q^2,J_1,J_2)$,
normalized over $d\zeta dq^2 dJ_1 dJ_2$,
has a fully factorized form in these variables
%
\begin{figure}
 {\includegraphics[height=.3\textheight]{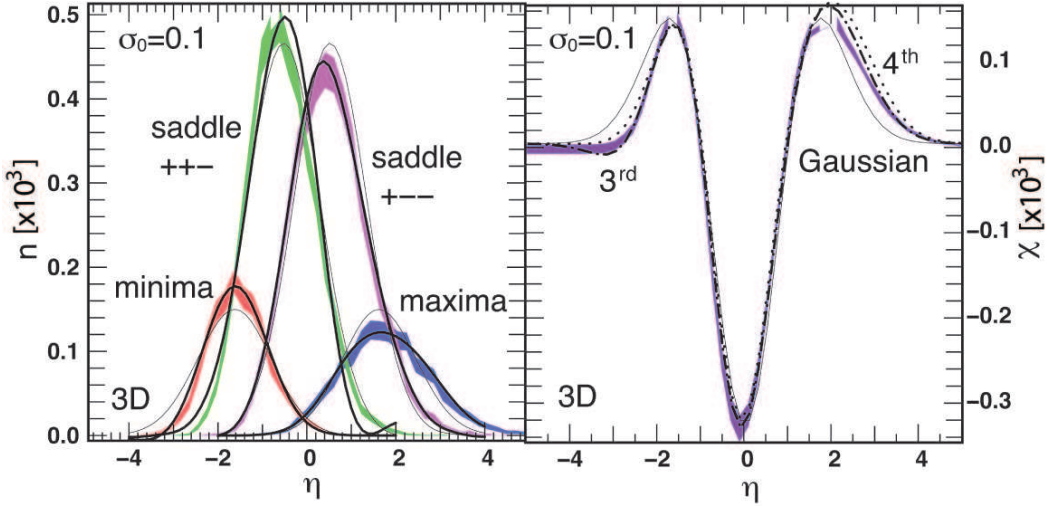}}
  \caption{The number of extrema ({\sl left}) and the Euler characteristic
({\sl right}) for  3D gravitational collapse. In dimensional units
both quantities are given per $R_*^3$ volume.
The shaded bands correspond to $2\sigma$ variations over the mean measurements
of 25 realizations, while the curves give the $3^{\rm rd}$ and
$4^{\rm th}$ order predictions. 
The Gaussian prediction is shown as a thin line. 
\label{fig:NG}
}
\end{figure}
Used as a kernel for the polynomial expansion, $G_{\rm 2D}$
leads to a non-Gaussian rotation invariant
JPDF in the form of the direct series in the products of
Hermite ($H_i$), for $\zeta$ and $J_1$, and Laguerre ($L_i$), for $q^2$
and $J_2$, polynomials:
\begin{equation}
P_{\rm 2D}(\zeta, q^2, J_1, J_2) =  G_{\rm 2D} \left[ 1 + 
\sum_{n=3}^\infty \sum_{i,j,k,l=0}^{i+2 j+k+2 l=n} 
\frac{(-1)^{j+l}}{i!\;j!\; k!\; l!} 
\left\langle \zeta^i {q^2}^j {J_1}^k {J_2}^l \right\rangle_c
H_i\left(\zeta\right) L_j\left(q^2\right)
H_k\left(J_1\right) L_l\left(J_2\right)  \cdots
\right]\,,
\label{eq:2DP_general}
\end{equation}
where $\sum_{i,j,k,l=0}^{i+2 j+k+2 l=n} $
stands for summation over all combinations
of non-negative $i,j,k,l$ such that $i+2j+k+2l$ adds 
to the order of the expansion term $n$.  A similar expression holds in 3D.
The coefficients of the expansion are ``centered'' moments, 
given by the differences of the actual moments and their Gaussian limit
(the latter vanishing when either $i$ or $k$ is odd):
$
\langle \zeta^i {q^2}^j {J_1}^k {J_2}^l\rangle_c =
\langle \zeta^i {q^2}^j {J_1}^k {J_2}^l\rangle
- (i-1)!! (k-1)!! j! l!\,.
$
Once the JPDF is known it is straightforward to compute any expectation of the field
which involve algebraic combinations of the invariants such as the  differential length,
${\partial {\cal L}^{\mathrm{ND}}}/{\partial \eta}$ (equation~(\ref{eq:NDdiff})), extrema counts, 
or the Euler characteristic. 
For instance,  the Euler characteristic can be computed completely (since there are no sign constraints on the eigenvalues of the Hessians)
by noting  that $I_N$ depends only linearly
on $J_{s \ge 2}$ (e.g., 
$ I_3 =  \left({J_1}^3 - 3 J_1 J_2 + 2 J_3 \right)/27
$
in 3D), hence all terms in JPDF of higher order in $J_{s \ge 2}$
do not contribute.  The 2D and 3D results 
can be combined in a very compact form if one re-expresses the 
``centered'' moments
back in terms of the field $\eta$ itself and the invariants $I_s$
\begin{eqnarray}
\lefteqn{\chi (\eta) = 
\frac{1}{2} \mathrm{Erfc} \left(\frac{\eta}{\sqrt{2}} \right)
\chi(-\infty) 
+ \frac{1}{\sqrt{2 \pi}} 
\exp\left(-\frac{\eta^2}{2}\right) \times
\frac{2}{(2 \pi)^{N/2}} \left(\frac{\gamma}{\sqrt{N}}\right)^N
\left[\vphantom{ \sum_{i,j,k=0}^{i+2 j+k=n} }
H_{N-1}(\eta) +
\right. } \nonumber \\
&& + \left. \sum_{n=3}^\infty  
\sum_{s=0}^N \gamma^{-s}
\sum_{i,j=0}^{i+2j=n-s}
\frac{(-N)^{j+s} (N-2) !! L_j^{(\frac{N-2}{2})}(0)}{i! (2j+N-2)!!} 
\left\langle \eta^i {q^2}^j I_s \right\rangle_c H_{i+N-s-1}(\eta)
\right]\,, \label{eq:euler}
\end{eqnarray}
where $i=0,s=N$ terms have been combined into the boundary term $\propto
\chi(-\infty)$ fixed by the topology of the manifold, and should
be omitted from the further sum.
Now if the departure from Gaussianity is induced by gravitationnal clustering, the cumulants occuring in equations~(\ref{eq:2DP_general})
and (\ref{eq:euler}) can be computed in the context of perturbation theory \cite{bernardeau,Matsubara0}, will scale like the growth factor $D(z)$, and
can be used to constrain the dark energy equation of state via
3D galactic surveys, or shed  light on the physics of the early Universe
through 2D CMB maps. Regarding the connectivity, equation~(\ref{eq:euler}) is illustrated  on figure~\ref{fig:NG} in this context.
In particular, the non-linear evolution of the number of saddle points and the number of peaks is accurately  predicted, which in turn should allow us to predict 
the mean degree, $\langle n \rangle$ of the peaks within each peak-patch given that each saddle-point connects to two peaks: $\langle n(z) \rangle= 2n_{\rm saddle}(z)/n_{\rm max}(z)$. When a non-Gaussian extension of equation~(\ref{eq:nB2D-2}) is derived
we should also be in a position to predict the non-linear evolution of the number of connecting streams on dark matter halos.

\bibliographystyle{aipproc}   

\bibliography{proceeding}
\end{document}